\documentclass[aps,prb,reprint]{revtex4-2}
\usepackage{amsmath,amssymb}
\usepackage{bm}

\usepackage{mathtools}

\usepackage{graphicx}
\usepackage{dcolumn}
\usepackage[dvipsnames]{xcolor}
\usepackage[caption=false,position=top,font=normalsize]{subfig}

\usepackage{braket}
\usepackage[dvips, bookmarkstype=toc, colorlinks=true, linkcolor=cyan, pdfborder={0 0 0}, bookmarks=true, bookmarksnumbered=true]{hyperref}

\usepackage{tikz}
\usepackage{ulem}

\newcommand*\YY[1]{\textcolor{red}{#1}}

\newcommand{\lr}[1]{\left({#1}\right)}

\newcommand{\lra}[1]{\left[{#1}\right]}

\let\origlim\lim
\renewcommand{\lim}[2]{\origlim\limits_{{#1}\rightarrow{#2}}}

\newcommand{\Sum}[2]{\sum\limits_{#1}^{#2}}

\newcommand{\eps}{\epsilon}

\newcommand{\ua}{\uparrow}
\newcommand{\da}{\downarrow}

\begin{document}

\title{Chiral superconducting diode effect by Dzyaloshinsky-Moriya interaction}

\author{Naratip Nunchot}
\author{Youichi Yanase}

\affiliation{Department of Physics, Graduate School of Science, Kyoto University, Kyoto 606-8502, Japan}

\date{\today}

\begin{abstract}

A two-component quasi-two-dimensional superconductor with Dzyaloshinsky-Moriya interaction is studied based on the Ginzburg-Landau and Bogoliubov-de Gennes theories. Under external in-plane magnetic fields, the order parameter of the superconducting state is a type of the Fulde-Ferrell state with a finite momentum of Cooper pairs due to the Dzyaloshinsky-Moriya interaction. It is shown that the superconducting diode effect can emerge when a supercurrent flows parallel to the external magnetic field, characteristic of chiral crystals. In the Bogoliubov-de Gennes theory, phase diagrams associated with the transition of the Cooper-pair momentum and the Josephson phase between spin-singlet and spin-triplet Cooper pairs are derived, and a close relationship with the diode quality factor is demonstrated. Implications of critical currents in the aspect of thermodynamics are also discussed. Based on such an argument, it is argued that the first-order phase transition in terms of Cooper-pair momentum and the coexistence of phases with different Cooper-pair momentum and Josephson phase can occur. The argument also implies the issue with the definition of critical currents calculated from the extremes of the supercurrent when metastable states exist. Comments on purely two-dimensional superconductors are also given.

\end{abstract}

\maketitle

\section{Introduction}

Nonreciprocal transport of matter has been researched from many aspects in recent years. Superconductors (SC) are one of the platforms for seeking such phenomena~\cite{Tokura}. To understand the nonreciprocal transport, starting with symmetry arguments can gain a general insight into the phenomena. According to the Onsager reciprocal relations~\cite{Onsager}, the necessary condition for such an effect is the breaking of the parity (P) and time-reversal (T) symmetries of the system. A class of superconductors that do not possess parity symmetry is classified as a noncentrosymmetric superconductor. A well-studied system of such a class is the Rashba superconductors with polar asymmetry in the crystalline structure. They can exhibit a nonreciprocal transport if the Zeeman coupling is considered. For convenience, we will term such systems the Rashba-Zeeman superconductors. The fluctuation current of the Rashba-Zeeman superconductors responding to an external electric field can become nonreciprocal \cite{Wakatsuki}. That means the transport coefficients such as paraconductivity also show nonreciprocity \cite{Wakatsuki, Hoshino}. In a purely two-dimensional SC, in which the resistive SC transition is the Berezinskii-Kosterlitz-Thouless (BKT) transition \cite{Kosterlitz, Nelson, Halperin}, resistivity also becomes nonreciprocal near the BKT transition in the low current limit \cite{Hoshino}. 

One of the interesting nonreciprocal effects in noncentrosymmetric superconductors is the superconducting diode (SD) effect. It refers to an effect that the magnitude of critical currents in the left-hand direction and the right-hand one is not equivalent. The SD effect has been observed in various experimental works \cite{Ando, Bauriedl, Lin, Narita}, and some theoretical formulations describing the effect have also been established \cite{Yuan, Daido1, Daido2, Scammell, He, Ilic}. The pictures of the mechanism of the SD effect are different depending on how superconductivity breaks down under electric currents. Upon increasing the external current, in an \textit{essentially} three-dimensional system, the Meissner phase can be destroyed by the growth of vortex loops \cite{Fisher}, and the vortex lattice phase, as well as vortex glass phase, can be destructed by the vortex depinning transition \cite{Nattermann}. Superconductivity can also be broken by the mechanism originating from the depairing of Cooper pairs \cite{Tinkham}. If the critical currents show nonreciprocity by the latter \cite{Yuan, Daido1, Daido2, Scammell, He, Ilic}, it is referred to as the \textit{intrinsic} SD effect \cite{Daido1, Daido2}. For example, the zero-field SD effect in the superconductor-ferromagnet heterostructures~\cite{Narita} is considered to be due to the intrinsic mechanism. 

We recall that the necessary condition for the SD effect is the broken P and T symmetry. Thus, time-reversal-symmetry-breaking fields such as external magnetic fields are needed. For example, the SD effect in Ref.~\cite{Ando} occurs due to the broken P symmetry in the crystalline structure and the broken T symmetry by the external magnetic fields. However, the SD effect can be possible even without an external magnetic field in twisted trilayer graphenes \cite{Lin}. This is considered to originate from a long-range order allowing time-reversal-symmetry-breaking fields. The theoretical formulation~\cite{Scammell} in such a case implies that the zero-field SD effect observed in Ref.~\cite{Lin} has essentially the same origin as the intrinsic SD effect studied in Refs.~\cite{Yuan, Daido1, He, Daido2, Ilic}, where the critical currents were calculated in the same way. Recently, the SD effect arising from dissipation, which causes time-reversal symmetry breaking in nonequilibrium steady states, has also been predicted~\cite{Daido_uni}.

In several theoretical works~\cite{Yuan, Daido1, Daido2, Ilic}, the intrinsic SD effect is studied with the Rashba-Zeeman superconductors. However, the P and T symmetries are also broken by the Dzyloshinzsky-Moriya (DM) interaction~\cite{Dzyal, Moriya} and Zeeman coupling. A phenomenological study on noncentrosymmetric superconductors such as a heavy-fermion superconductor $\mathrm{CePt_3Si}$ \cite{Frigeri} points out that mixed singlet-triplet pairing interactions giving rise to parity-mixed order parameters originate from the DM interaction. Moreover, a weak coupling theory of the mixed singlet-triplet pairing interaction is also developed to study the static spin susceptibility of the Fermi liquid. In Ref. \cite{Samo}, such a theory is constructed to study the gap structure in noncentrosymmetric superconductors such as cubic compounds $\mathrm{Li_2(Pd_{1-x}Pt_x)_3B}$. There is also a study on a non-Hermitian system \cite{Ghatak}, in which the theory of superconductivity is developed by dropping out the assumption on the hermiticity of the Hamiltonian. In such a case, anti-Hermitian pairing can emerge from the DM interaction.  

In this paper, we study a superconductor with the Zeeman coupling and DM interaction under external in-plane magnetic fields. Because of mixed singlet-triplet pairing interactions, the system becomes a multi-component superconductor. We simplify the problem by choosing the asymmetric axis of the DM interaction in one direction so that the model becomes a two-component superconductor with a non-vanishing internal Josephson coupling, which is a consequence of the DM interaction and will be given in the main text. The external magnetic field directions, together with the supercurrent directions, will be chosen to be parallel to that axis. In this way, the model is essentially different from those in Refs.~\cite{Yuan, Daido1, Daido2, He, Ilic}, where the external magnetic field and supercurrent are perpendicular to each other. In the following, we will show that our model can exhibit the intrinsic SD effect in both the Ginzburg-Landau (GL) and Bogoliubov-de Gennes (BdG) frameworks, and obtain the phase diagrams. In the Rashba-Zeeman and Ising superconductors, the SD effect does not occur in this parallel geometry. Importantly, the vortex drift motion by applied supercurrent is suppressed in this setup, making it easier to observe the intrinsic SD effect. Such an SD effect in parallel geometry can occur in systems with chiral asymmetry, and thus, we call it the chiral SD effect. Moreover, we will go deep into how a critical current is calculated in an aspect of thermodynamics and show that there are some problems with the definition of critical currents caused by the current mean-field framework. Lastly, we will give some comments related to the SD effect in a purely two-dimensional (2D) system. 
 
The paper is organized as follows: In Sec.~II, we describe our model. In Sec.~III, the GL mean field theory of the model is constructed first, and we then discuss what the critical currents imply based on a thermodynamics aspect before reporting a numerical result of the model in the counterpart framework. Section~IV is devoted to the BdG mean field theory of the model. There, the gap equations, the formula of free energy, and the parameters for the numerical calculation are discussed first. Then, several numerical results are shown. At the end of Sec.~IV, we apply the theory for critical currents developed in Sec.~III to the model and discuss the consequence of it. In Sec.~V, we give two comments on a purely 2D system. One is related to the BKT transition, and the other is about the absence of the SD effect in a strict sense. We end the paper with a summary and discussions. Natural units $\hbar = k_{\mathrm{B}}=c=1$ will be used in this paper.

\section{Model}
\label{sec:model}

We consider a thin superconducting film lying on the $z$-$x$ plane under in-plane magnetic fields $\bm{h} = h\bm{e}_z$, where $h=\mu_B H$. Such a film is treated here as a 2D-like system whose picture of the superconducting transition does not differ from a 3D counterpart. The partition function of the system is given below:
\[   Z = \mathrm{Tr}\, e^{-S},  \tag{1}\label{eq1}  \]
\[ S = \int_0^{1/T}d\tau\, \lra{ \Sum{\bm{q}}{} \bar{a}_{\alpha}(\tau,\bm{k}) \partial_{\tau} a_{\alpha}(\tau,\bm{k}) + \mathcal{H}(\bar{a},a) }.  \tag{2}  \]
Here, $a_{\alpha}$ and $\bar{a}_{\alpha}$ are Grassmann variables corresponding to the annihilation and creation operators of fermions with spin $\alpha= \ua,\da$ and 2D momentum $\bm{k} =(k_1,k_3)$, where the subscripts $i=1,2,3$ represent $x,y,z$-axis, respectively. $\tau$ denotes imaginary time and $T$ is the temperature of the system. The Hamiltonian $\mathcal{H}$ is given by $\mathcal{H}=\mathcal{H}_0+\mathcal{H}_{\mathrm{int}}$, where $\mathcal{H}_0$ is the free part and $\mathcal{H}_{\mathrm{int}}$ is the interaction part. Each of them is given as follows:
\begin{align*} 
\mathcal{H}_0 = \Sum{\bm{k}}{} \lr{ \eps_{\bm{k}}\delta_{\alpha\beta} + \bm{h}\cdot\bm{\sigma}_{\alpha\beta} } \bar{a}_{\alpha}(\tau,\bm{k})a_{\beta}(\tau,\bm{k}),  \tag{3a}
\end{align*} 
\begin{align*}     
 \mathcal{H}_{\mathrm{int}} = - \frac{1}{4} \sideset{}{'}\sum_{\bm{k},\bm{k}',\bm{q}} & V_{\alpha\beta\gamma\delta}(\bm{k},\bm{k}')  \bar{a}_{\alpha}(\tau,-\bm{k} \! +\! \bm{q})\bar{a}_{\beta}(\tau,\bm{k}) \\
 &\times a_{\gamma}(\tau,\bm{k}')a_{\delta}(\tau,-\bm{k}'\! + \!\bm{q}),  \tag{3b} \label{eq3b}
\end{align*}  
with 
\begin{align*} 
& \hspace{5mm} V_{\alpha\beta\gamma\delta}(\bm{k},\bm{k}')= \\
& \hspace{12mm} g\,(i\sigma_2)^{\dagger}_{\alpha\beta}(i\sigma_2)_{\gamma\delta} \! +   v_a \,(i\hat{k}_3\sigma_3\sigma_2)^{\dagger}_{\alpha\beta} (i\hat{k}'_3\sigma_3\sigma_2)_{\gamma\delta} \\
& \hspace{12mm} + \! v_m \big[ \,(i\hat{k}_3\sigma_3\sigma_2)^{\dagger}_{\alpha\beta} (i\sigma_2)_{\gamma\delta} \!+ \! (i\sigma_2)^{\dagger}_{\alpha\beta} (i\hat{k}'_3\sigma_3\sigma_2)_{\gamma\delta} \big],     \tag{3c}   \label{eq3c}   
\end{align*}
where $\eps_{\bm{k}}=\bm{k}^2/2m-\mu$, $\mu=p^2_{\rm F}/2m$ is the Fermi energy, and $[\hat{\xi}(\bm{k})]_{\alpha\beta} \equiv \xi_{\alpha\beta}(\bm{k}) = \eps_{\bm{k}}\delta_{\alpha\beta} + h\,(\sigma_{3})_{\alpha\beta}$. $\sigma_3$ denotes the $z$-component of the Pauli matrices, where $(\sigma_3)_{\ua\ua}=- (\sigma_3)_{\da\da}  =1$ and $(\sigma_3)_{\ua\da}=(\sigma_3)_{\da\ua}  =0$. $\hat{k}_i$ denotes the cosine direction of $\bm{k}$ along $i$-axis, defined as $\hat{k}_i \equiv \bm{e}_i\cdot\bm{k}/|\bm{k}|$, where $\bm{e}_1 = \bm{e}_x$ and $\bm{e}_3=\bm{e}_z$. The first and second terms on the right-hand side of Eq.~(\ref{eq3c}) represent the $s$-wave interaction with coupling strength $g$ and the $p$-wave interaction with coupling strength $v_a$, respectively. The terms in the third line of Eq.~(\ref{eq3c}) represent the DM interaction with coupling strength $v_m$ following the form obtained in Ref.~\cite{Samo}. Such a form comes from the original form of the DM interaction: $\mathcal{H}_{\mathrm{DM}} = i\sum_{\bm{q}}\bm{V}_{m}(\bm{q})\cdot(\bm{S}_{\bm{ q}}(\tau)\times\bm{S}_{-\bm{q}}(\tau) )$, where $\bm{S}_{\bm{q}}(\tau)=\sum_{\bm{k}}\bm{\sigma}_{\alpha\beta} \, \bar{a}_{\alpha}(\tau,\bm{k})a_{\beta}(\tau,\bm{k}\!+\!\bm{q})$. In the considered model, $\bm{V}_m(q)$ is chosen to be proportional to $v_m\hat{q}_3\bm{e}_z$. The prime symbol on the summation notation in Eq.~(\ref{eq3b}) means that the summation of $\bm{q}$ is restricted for some values, and the summation of $\bm{k}$ and $\bm{k}'$ is performed on the range $[\Lambda_-,\Lambda_+]$, where $\Lambda_+> p_{\rm F} > \Lambda_->0$. The width $\Lambda \equiv \Lambda_+-\Lambda_-$, which corresponds to the energy width $2\omega_{\rm c}$,  will be assumed hereafter so that $\Lambda \gg \mathrm{max}\, |\bm{q}|$.

By symmetry of the system, the Cooper-pair momentum $\bm{q}$, with which the superconducting state is stable, should have only a $z$-axis component. Therefore, in the following, we will only focus on the case where $\bm{q}$ lies on the $z$-axis. Then we have $\bm{q}=q\bm{e}_z$. This also implies the situation where the supercurrent flows only along the $z$-axis. In the setup discussed in the preceding paragraph, the external magnetic fields and asymmetric axis of the DM interaction are also on the $z$-axis. Accordingly, hereafter, we will only consider the case when the supercurrent flow is parallel to the external magnetic fields and asymmetric axis of the DM interaction.

It is much more convenient to replace $\bm{k}$ and $\bm{k}'$ in Eqs.~(\ref{eq3b}) and (\ref{eq3c}) with $\bm{k}+\bm{q}/2$ and $\bm{k}' +\bm{q}/2$, respectively. After taking the procedure, $\hat{k}_3$ becomes, up to $q^2$-order,
\[     \hat{k}_3 \rightarrow \hat{k}_3 + \frac{1}{2} \hat{k}^2_1 \lr{\frac{q}{k}} - \frac{3}{8} \hat{k}^2_3 \hat{k}^2_1 \lr{\frac{q}{k}}^2.  \tag{4}   \]
In this paper, we assume $|q|\ll p_{\rm F}$. This allows us to neglect all terms depending on $q$ in the above equation. From this assumption and Eqs.~(\ref{eq3b}) and (\ref{eq3c}), one can obtain the following:
\begin{align*}     
\mathcal{H}_{\mathrm{int}} = - \frac{1}{4} \sideset{}{'}\sum_{\bm{k},\bm{k}',\bm{q}} & V_{\alpha\beta\gamma\delta}(\bm{k},\bm{k}')  \bar{a}_{\alpha}(\tau,-\bm{k}_-)\bar{a}_{\beta}(\tau,\bm{k}_+) \\
&\times a_{\gamma}(\tau,\bm{k}'_+)a_{\delta}(\tau,-\bm{k}'_-),  \tag{5a} \label{eq5a}
\end{align*}
\begin{align*} 
& \hspace{5mm} V_{\alpha\beta\gamma\delta}(\bm{k},\bm{k}') \\
& \hspace{12mm} \simeq g\,(i\sigma_2)^{\dagger}_{\alpha\beta}(i\sigma_2)_{\gamma\delta} \! +  v_a \,(i\hat{k}_3\sigma_3\sigma_2)^{\dagger}_{\alpha\beta} (i\hat{k}'_3\sigma_3\sigma_2)_{\gamma\delta} \\
& \hspace{12mm} + v_m \big[ \,(i\hat{k}_3\sigma_3\sigma_2)^{\dagger}_{\alpha\beta} (i\sigma_2)_{\gamma\delta} \! + \! (i\sigma_2)^{\dagger}_{\alpha\beta} (i\hat{k}'_3\sigma_3\sigma_2)_{\gamma\delta} \big],     \tag{5b} \label{eq5b}      
\end{align*}
where $\bm{k}_{\pm}=\bm{k}\pm\bm{q}/2$ and $\bm{k}'_{\pm}=\bm{k}'\pm\bm{q}/2$. 

Next, let us consider the diagonalization of Eq.~(\ref{eq5b}) and discuss its eigenvalues. The result of diagonalization is in the form $\bm{\Psi}^{\dagger}_{\alpha\beta\bm{k}}\Lambda\bm{\Psi}_{\gamma\delta\bm{k}'}$, where
\[   \Lambda = \mathrm{diag}\lr{\lambda_+,\lambda_-}, \tag{6}\]   
\[\lambda_{\pm} = \frac{1}{2}\lra{ g+v_a \pm \sqrt{(g+v_a)^2+4 \, (v^2_m-gv_a) }},   \tag{7} \label{eq7} \]
\begin{align*}
 \bm{\Psi}_{\alpha\beta\bm{k}} & \equiv \begin{pmatrix}
\Psi_{1\alpha\beta\bm{k}} \\
\Psi_{2\alpha\beta\bm{k}}
\end{pmatrix} \\
&=  \begin{pmatrix}
\cos\chi \, (i\sigma_2)_{\alpha\beta} + \sin\chi \, (i\hat{k}_3\sigma_3\sigma_2)_{\alpha\beta}  \\
\sin\chi \, (i\sigma_2)_{\alpha\beta} -\cos\chi \, (i\hat{k}_3\sigma_3\sigma_2)_{\alpha\beta}  \tag{8a}
\end{pmatrix},
\end{align*}
and
\[\tan\chi =(v_a-\lambda_-)/v_m. \tag{8b} \] 
Now, let us look at the sign of $\lambda_-$. From Eq.~(\ref{eq7}), it is apparent that if $v_m>\sqrt{gv_a}$, then $\lambda_-$ becomes negative. In that case, the terms associated with $\lambda_-$ do not participate in ordering the superconducting state because the interaction between fermions through this mode is effectively repulsive. Therefore, the system is reduced to a one-component order parameter system, and, in the Hamiltonian, the term associated with $\lambda_-$  is responsible for a correction term of the free part $\mathcal{H}_0$. A detailed study of such a case will be our future work. In this paper, we consider the case with positive $\lambda_-$. In this case, Eq.~(\ref{eq5b}) becomes 
\[        V_{\alpha\beta\gamma\delta} (\bm{k},\bm{k}') = \lambda_+ \Psi^{\dagger}_{1\alpha\beta\bm{k}} \Psi_{1\gamma\delta\bm{k}'}  +\lambda_- \Psi^{\dagger}_{2\alpha\beta\bm{k}} \Psi_{2\gamma\delta\bm{k}'}~.  \tag{9} \label{eq9} \]
As we will demonstrate in a moment below, the presence of two $\Psi_{i\alpha\beta\bm{k}}$'s in Eq.~(\ref{eq9}) means that the system is described by two parity-mixed order parameters responsible for ordering the superconducting state. 

%Let $\bm{\psi}_{\alpha\beta\bm{k}} \equiv \lra{ (i\sigma_2)_{\alpha\beta} , (i\hat{k}_3\sigma_3\sigma_2)_{\alpha\beta} }$, we can rewrite $V_{\alpha\beta\gamma\delta}(\bm{k},\bm{k}')$ to be in the matrix form as below: 
%\[ V_{\alpha\beta\gamma\delta}(\bm{k},\bm{k}') =\bm{\psi}^{\dagger}_{\alpha\beta\bm{k}} \begin{pmatrix}
%g & v_m \\
%v_m & v_a 
%\end{pmatrix} \bm{\psi}_{\gamma\delta\bm{k}'} ~.      \tag{6}   \]

In the last of this section, we derive the effective action of the model. By introducing auxiliary boson fields, $\{ \varphi_i(\bm{q}),\bar{\varphi}_i(\bm{q}) \}~(i=1,2)$, and using the Hubbard-Stratonovich transformation, Eq.~(\ref{eq5a}) is decoupled. At this time, the partition function becomes 
\[    Z = \mathrm{Tr}_{\varphi_{i},a_{\alpha}} e^{- \int_0^{1/T}d\tau\, H' }, \tag{10} \label{eq10} \]
where the subscript of $\mathrm{Tr}$ means the integral with respect to $\varphi_i, \bar{\varphi}_i, a_{\alpha}$ and $\bar{a}_{\alpha}$. Here, $H'$ is given below:
\[    H' = H_{\mathrm{BdG}} + \sideset{}{'}\sum_{\bm{q},i} \frac{1}{|\lambda_{i}|}\bar{\varphi}_i(\tau,\bm{q})\varphi_i(\tau,\bm{q}),   \tag{11}     \]
where
\begin{align*}   
H_{\mathrm{BdG}} = & \sideset{}{'}\sum_{\bm{q}} \bar{a}_{\alpha}(\tau,\bm{q}) \partial_{\tau} a_{\alpha}(\tau,\bm{q}) + H_0 \\
&+\frac{1}{2} \, \sideset{}{'}\sum_{\bm{q},\bm{k}}  \Delta_{+\alpha\beta}(\tau,\bm{q},\bm{k})  a_{\alpha}(\tau,\bm{k}_+)a_{\beta}(\tau,-\bm{k}_-) \\ 
&+\frac{1}{2} \, \sideset{}{'}\sum_{\bm{q},\bm{k}} \Delta_{-\alpha\beta}(\tau,\bm{q},\bm{k}) \bar{a}_{\alpha}(\tau,-\bm{k}_-)\bar{a}_{\beta}(\tau,\bm{k}_+), \tag{12} \label{eq12}
\end{align*}
and
\begin{align*}   
\hat{\Delta}_+(\tau,\bm{q},\bm{k})&= \bar{\varphi}_1(\tau,\bm{q}) \hat{\Psi}_{1\bm{k}} + \bar{\varphi}_2(\tau,\bm{q}) \hat{\Psi}_{2\bm{k}}, \tag{13a} \\ 
\hat{\Delta}_-(\tau,\bm{q},\bm{k})&=\varphi_1(\tau,\bm{q}) \hat{\Psi}^{\dagger}_{1\bm{k}} + \varphi_2(\tau,\bm{q}) \hat{\Psi}^{\dagger}_{2\bm{k}}, \tag{13b}
\end{align*}
are the so-called pair fields. In the above equation, we use the hat symbol `$\hat{[\cdot]}$' to represent each variable in a compact matrix form. For example, $[\hat{\Psi}_{i\bm{k}}]_{\alpha\beta}=\Psi_{i\alpha\beta\bm{k}}$. The dagger denotes the Hermitian conjugate. By introducing four-component Nambu fields:
\begin{widetext}

\begin{align*}
&\Phi(\tau,\bm{k}) \equiv \frac{1}{\sqrt{2}} \lra{ a_{\ua}(\tau,\bm{k}_+) , a_{\da}(\tau,\bm{k}_+) ,\bar{a}_{\ua}(\tau,-\bm{k}_-) , \bar{a}_{\da}(\tau,-\bm{k}_-)}^{\mathrm{T}},     \tag{14a} \\
&\bar{\Phi}(\tau,\bm{k}) \equiv \frac{1}{\sqrt{2}} \lra{ \bar{a}_{\ua}(\tau,\bm{k}_+) , \bar{a}_{\da}(\tau,\bm{k}_+) ,a_{\ua}(\tau,-\bm{k}_-) , a_{\da}(\tau,-\bm{k}_-)},  \tag{14b}
\end{align*}
Eq.~(\ref{eq12}) can be rewritten as follows:
\begin{align*}
H_{\mathrm{BdG}} &= \sideset{}{'}\sum_{\bm{k}} \bar{\Phi}(\tau,\bm{k}) \, \mathcal{G}^{-1}(\tau,\bm{q},\bm{k}) \, \Phi(\tau,\bm{k}) +  \frac{1}{2}\sideset{}{'}\sum_{\bm{k},\bm{q}'\not=\bm{q}} \bar{\Phi}(\tau,\bm{k}) \, \mathcal{G}^{-1}_{\mathrm{cor}}(\tau,\bm{q}',\bm{k}) \, \Phi(\tau,\bm{k}),       \tag{15} \label{eq15}                   \end{align*}

\end{widetext}
where
\[ \mathcal{G}^{-1}(\tau,\bm{q},\bm{k}) \equiv \begin{pmatrix}
\partial_{\tau} + \hat{\xi} (\bm{k}_+) & \hat{\Delta}_-(\tau,\bm{q},\bm{k}) \\
\hat{\Delta}_+(\tau,\bm{q},\bm{k}) & \partial_{\tau} - \hat{\xi} (-\bm{k}_-)
\end{pmatrix}, \label{16a}  \tag{16a} \]
and
\[ \mathcal{G}^{-1}_{\mathrm{cor}}(\tau,\bm{q}',\bm{k}) \equiv \begin{pmatrix}
0 & \hat{\Delta}_-(\tau,\bm{q}',\bm{k}) \\
\hat{\Delta}_+(\tau,\bm{q}',\bm{k}) & 0
\end{pmatrix}.   \tag{16b}   \]
Hereafter, we assume that the main contribution of Eq.~(\ref{eq10}) is only the first term of the right-hand side of Eq.~(\ref{eq15}). Therefore, the remaining term in that equation will be neglected. This treatment means we are considering the Fulde-Ferrell (FF) state of superconductors~\cite{Fulde}. In this FF state, the order parameters in a real-space representation, namely $\varphi_i(z)$, are proportional to $\exp\,(iqz)$, and the amplitude of the gap function is spatially homogeneous. After integrating Eq.~(\ref{eq10}) over the fermion fields, we obtain the effective action $S_{\mathrm{eff}}$ as follows:
\begin{align*}  
 S_{\mathrm{eff}}[\bar{\varphi}_i,\varphi_i] \simeq &  \int_0^{1/T}d\tau\,  \Big[ -\frac{1}{2}\sideset{}{'}\sum_{\bm{k}}  \mathrm{Tr}  \ln \mathcal{G}^{-1}(\tau,\bm{q},\bm{k})  \\
 &+ \Sum{i}{}\frac{1}{|\lambda_{i}|}\bar{\varphi}_i(\tau,\bm{q})\varphi_i(\tau,\bm{q}) \Big],  \tag{17}   \label{eq17}
 \end{align*}
where $\mathrm{Tr}$ is a trace over eigenvalues of $\mathcal{G}^{-1}(\tau,\bm{q},\bm{k})$. The approximation notation is used above because we keep the range of the summation of momentum ${\bm k}$ unchanged from Eq.~(\ref{eq3b}). Equation~(\ref{eq17}) becomes exact when $q=0$. The above equation should be a good approximation if $|q|\ll\Lambda$. Below, we will neglect the dependence of $\varphi_i$ on boson Matsubara frequencies.

\section{Ginzburg-Landau Theory}
\label{sec:GL}

In this section, we study the effective action $S_{\mathrm{eff}}$ derived in the previous section in the GL theory framework. After deriving the GL free energy, we then discuss critical supercurrent and show a numerical calculation of the model in the GL framework. 
 
 \subsection{GL free energy}

To derive the GL free energy, we expand $S_{\mathrm{eff}}$ to the fourth-order terms of the order parameters. Following a procedure in Ref.~\cite{Simons}, we first write $\mathcal{G}^{-1}$ in the form $\mathcal{G}^{-1}=\mathcal{G}^{-1}_0 +\hat{\Delta}$, where 
 \begin{align*}  
 \mathcal{G}^{-1}_0(\omega_n,\bm{q},\bm{k}) &\equiv \begin{pmatrix}
-i \omega_n \!+  \hat{\xi} (\bm{k}_+) & 0 \\
0 & - i\omega_n \! - \hat{\xi}(-\bm{k}_-)
\end{pmatrix}, \tag{18a} \label{eq18a} \\
\hat{\Delta}(\bm{q},\bm{k}) &\equiv \begin{pmatrix}
0 & \Delta_-(\bm{q},\bm{k}) \\
\Delta_+(\bm{q},\bm{k}) & 0
\end{pmatrix},   \tag{18b} \label{eq18b}  
\end{align*} 
and $\omega_n$ are fermion Matsubara frequencies. Then, by using the formula $\ln \, [\mathcal{G}^{-1}_0(1+\mathcal{G}_0\hat{\Delta})] = \ln \mathcal{G}^{-1}_0 -\sum^{\infty}_{n=1}\frac{(-1)^n}{n}[\mathcal{G}_0\hat{\Delta}]^n$, we obtain the GL free energy $F_{\rm GL}$ as below:
\begin{align*}
F_{\rm GL} = \alpha_1\varphi^2_1&+\alpha_2\varphi^2_2+2\alpha_3\varphi_1\varphi_2\cos\phi \\
&+\beta_1\varphi^4_1+\beta_2\varphi^4_2 \\
&+2\varphi_1\varphi_2 \lr{\beta_3\varphi^2_1 + \beta_4\varphi^2_2} \cos\phi \\
&+ \varphi^2_1\varphi^2_2 \lr{\beta_5 + 4\beta_6\cos^2\phi},  \tag{19} \label{eq19}
\end{align*}
where $\varphi_i$ here is the magnitude of the auxiliary boson fields satisfying $\partial F_{\mathrm{GL}}/\partial \varphi_i=\partial F_{\mathrm{GL}}/\partial \phi=0$, and $\phi$ is the phase difference, or the so-called internal Josephson phase, between $\varphi_1$ and $\varphi_2$. The free energy of the normal state, coming from the contribution $\ln\mathcal{G}^{-1}_0$, had been subtracted already in the above equation. Coefficients $\alpha_i$ are calculated up to $q^2$-order and $h$-linear order, while coefficients $\beta_i$ are evaluated at $O(1)$. For a $q^n$-order term with $n\geq 1$, the cutoff momentum are $\Lambda_-=0$ and $\Lambda_+=\infty$. For a $O(1)$ term, $\Lambda_-=0$ but $\Lambda_+$ is given to be the momentum corresponding to the energy $\omega_{\rm c}$. See Appendix A for the explicit expression of each coefficient. From the above, the GL free energy $F_{\rm GL}$ has trivial stationary points at $\cos\phi = \pm 1$. A nontrivial one, where $\cos\phi\not=\pm1$, is also possible and given below:
\[ \cos\phi = - \frac{1}{4\varphi_1\varphi_2\beta_6}\lr{ \alpha_3 + \beta_3\varphi^2_1 + \beta_4\varphi^2_2 }.   \tag{20}  \]
However, near the SC phase transition under a low magnetic field, since the amplitude of order parameters is small, no nontrivial solution satisfies the above equation. Even at a relatively high magnetic field, where the GL theory in the context of this paper becomes invalid, numerical results show that such solutions tend to be metastable in the range examined. Therefore, we ignore possible nontrivial solutions in the following.

\subsection{Critical current}

Next, let us discuss critical currents. In literature such as Ref.~\cite{Daido1}, it is shown that, based on a mean-field microscopic theory, a supercurrent with Cooper-pair momentum $\bm{q}$, $\bm{I}(\bm{q})$, is given by
 \[       \bm{I}(\bm{q}) = \frac{\partial F(\bm{q})}{\partial \bm{q}}, \tag{21}   \label{eq21}  \]
where $F(\bm{q})$ is the Helmholtz free energy or condensation energy such as given by Eq.~(\ref{eq19}) or (\ref{eq25}). Here, we absorb all unnecessary coefficients into the definition of the current $\bm{I}(\bm{q})$. The critical current in a given direction is determined by finding the maximum value of supercurrent magnitude in that direction. As aforementioned in the last section, here we focus on the case where the supercurrent flows in the $z$-axis direction. Hereafter, let $I_+(q)$ and $I_-(q)$ be supercurrents flowing in the positive and negative $z$-axis direction, respectively. Then, the critical currents in the positive $z$-axis direction, $I_{+{\rm c}}$, and in the negative $z$-axis direction, $I_{-\rm{c}}$, are given by
\[             I_{\pm {\rm c}}  =   \mathrm{max}_{q} \,|I_{\pm}(q) |.   \tag{22} \label{eq22} \]
In this way, the diode quality factor $r_{\rm d}$ is defined as below:
\[     r_{\rm d} = \frac{I_{+{\rm c}}-I_{-{\rm c}}}{I_{+\rm{c}}+I_{-\rm{c}}}. \tag{23} \label{23}     \]
Equation~(\ref{eq22}) will be evaluated intensely in the next subsection and Sec.~\ref{sec:BdG}.

Now, let us consider the meaning of the critical current from a thermodynamic point of view. From Eq.~(\ref{eq21}), one may regard that a critical current is obtained by minimizing $G(\bm{q},\bm{I}) = F(\bm{q}) - \bm{I}\cdot\bm{q}$. The minimum value of $G(\bm{q},\bm{I})$ is the Gibbs free energy $G(\bm{I})=\mathrm{min}_{\bm{q}} \, \{ G(\bm{q},\bm{I}) \} $. Indeed, there is prior research \cite{Samotr} asserting that the system should realize in the state obtained by minimizing $G(\bm{q},\bm{I})$. As clear from the definition of $G(\bm{I})$, if the state is thermodynamically stable in the above sense, then $G(\bm{I})$ is concave with respect to $\bm{I}$. By this consideration, a critical current can be interpreted as a supercurrent in which $G(\bm{I})$ becomes convex with respect to $\bm{I}$ indicating thermodynamical instability. This interpretation can give a critical current different from Eq.~(\ref{eq22}) when multiple metastable states appear. Further details will be discussed in Sec.~\ref{sec:phase_coexistence}.

In the previous paragraph, we discussed critical currents from a thermodynamic aspect by considering the supercurrent and Cooper-pair momentum as thermodynamic variables like pressure or an external magnetic field. As a result, the framework is the same as equilibrium thermodynamics although there is a constant external current flowing as a supercurrent in the system. Moreover, it also means that $\bm{q}$ is constant in time by itself. It has to be examined further from both theoretical and experimental sides whether $\bm{q}$ can still be constant in time under an external current.

\subsection{Numerical results of GL model}
\label{sec:numerical_GL}

\begin{figure}[b]

\captionsetup{font=normal,justification=raggedright,singlelinecheck=false}

\begin{minipage}{\hsize}
\subfloat[\label{fa}]{\includegraphics[width=7.2cm]{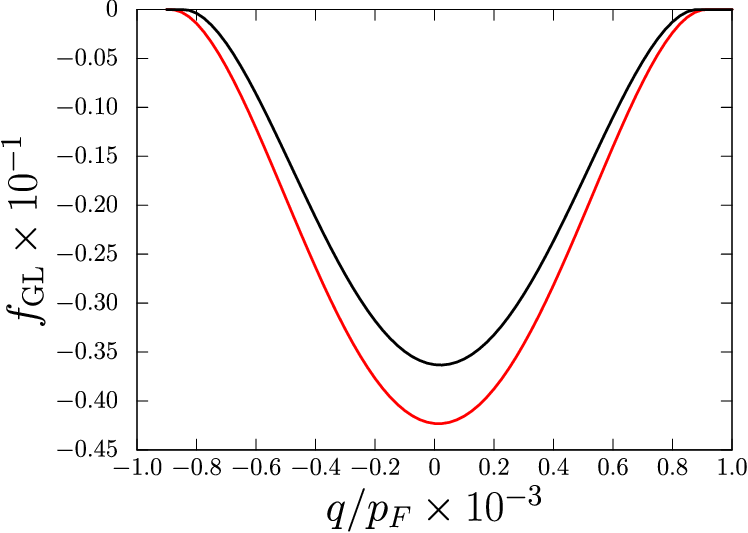}}
\end{minipage}

\begin{minipage}{\hsize}
\subfloat[\label{jb}]{\includegraphics[width=7.2cm]{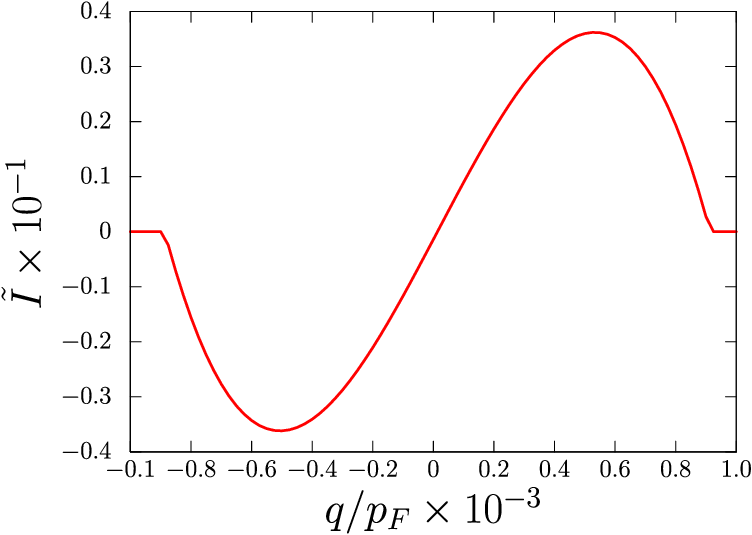}}
\end{minipage}

\caption{Cooper-pair momentum dependences of (a) the reduced dimensionless mean-field free energy $f_{\mathrm{GL}}$ and (b) corresponding supercurrent $\tilde{I}$ at temperature $t=0.9$ and magnetic field $\tilde{h}=0.1$. In panel (a), the red and black lines show the GL free energy with the internal Josephson phase $\phi=0$ and $\pi$, respectively. In panel (b), the supercurrent in the stable state (red line in panel (a) for $\phi=0$) 
is plotted.}

\label{fjg}
\end{figure}

Here, we point out a feature of the GL model in Eq.~(\ref{eq19}) that differs from a single-component superconductor studied in Refs.~\cite{Daido1, Daido2, Ilic}. There, it is shown that, for a one-component order parameter of superconductivity, the quadratic term such as $\alpha_1$ must be calculated up to $q^3$-order to realize the SD effect. In considering a system with two-component order parameters, however, stopping at $q^2$-order of each $\alpha_i$ term is sufficient to realize the SD effect. We can verify this with explicit calculations.  Hereafter, we choose $gN(0)=0.2,~v_aN(0)=0.1$,~$v_mN(0)=0.01$,~$\mu/T_{\rm{c}0}=10^{3}$, and $\omega_{\rm c}/\mu=0.13$, where $T_{\rm{c}0}$ is the zero-field SC transition temperature and $N(0)$ is the density of states at the Fermi level (see Appendix A). Indeed, it is not necessary to specify the last two parameters when working in the GL framework. However, for convenience in the numerical calculation and to compare the results with the BdG theory in the next section, we assign those values here as well. Substituting the minimized free energy of Eq.~(\ref{eq19}) for each $q$ into Eq.~(\ref{eq21}), the supercurrent with Cooper-pair momentum $q$ is computed. Here we define the reduced dimensionless GL free energy $f_{\mathrm{GL}}$ and supercurrent $\tilde{I}$ as $f_{\mathrm{GL}} \equiv F_{\mathrm{GL}}(N(0)T^2_{{\rm c}0})^{-1}$ and $\tilde{I} \equiv 2 I (v_F N(0)T_{{\rm c}0})^{-1}$, respectively. The dimensionless free energy is a function of the normalized Cooper-pair momentum $q/p_{\rm F}$, normalized temperature $t\equiv T/T_{{\rm c}0}$, and normalized external magnetic field $\tilde{h}\equiv h/T_{{\rm c}0}=\mu_{\rm B} H/T_{{\rm c}0}$. We show the $q$-dependences of $f_{\mathrm{GL}}$ and $\tilde{I}$ of the GL model in Eq.~(\ref{eq19}) at normalized temperature $t=0.9$ and normalized magnetic field $\tilde{h}=0.1$ in Fig.~\ref{fjg}. As seen in Fig.~\ref{fa}, the GL free energy with the Josephson phase $\phi=0$ is smaller than that with $\phi=\pi$. This result is consistent with the BdG mean-field theory in the next section. Moreover, despite being difficult to see with the naked eye, the free energy takes the global minimum at finite $q$ with $|q| \ll p_{\rm F}$, and the critical supercurrents in positive and negative directions indeed differ in magnitude, giving the diode quality factor $r_{\rm dGL}=4.5\times10^{-4}$. In this manner, the SD effect in the two-component superconductors can be realized without $q^3$-order terms in $\alpha_i$, in contrast to the single-component superconductors. 

To make this result intuitive, let us solve a couple of the GL equations associated with Eq.~(\ref{eq19}) at $\phi=0$. Obtaining the exact solution is extremely difficult because the equations are cubic. For simplicity, let us consider the case with $\varphi_1\gg \varphi_2$. This assumption is appropriate for the parameters considered in this section. With this approximation, $\varphi_2$ can be expressed as a linear term of $\varphi_1$, namely $\varphi_2\simeq\alpha_3\varphi_1/\alpha_2$, and $\varphi^2_1$ can be solved to the first order of $\alpha_3/\alpha_2$ (see Appendix B). Substituting the expression of $\varphi^2_1$ into Eq.~(\ref{eq19}), one can further evaluate $F_{\mathrm{GL}}$ to the first order of $\alpha_3/\alpha_2$ (see Appendix B). From the asymptotic equation of $F_{\mathrm{GL}}$, Eq.~(\ref{eqA12}), it becomes clear that if one rewrites $F_{\mathrm{GL}}$ in the form $-\tilde{\alpha}^2/{4\tilde{\beta}}$, $q^n$-order terms where $n\geq 3$ are induced in $\tilde{\alpha}$, signaling the intrinsic SD effect. However, further numerical calculation shows that Eq.~(\ref{eqA12}) gives a diode quality factor $\tilde{r}_{\textrm{dGL}} =-1.7\times10^{-4}$ at $t=0.9$ and $\tilde{h}=0.1$, incompatible with the estimation $r_{\rm dGL}=4.5\times10^{-4}$ in the previous paragraph. The deviation from $r_{\textrm{dGL}}$ is due to ignoring the higher order terms of $\alpha_3/\alpha_2$ in Eq.~(\ref{eqA12}). Indeed, if one substitutes Eq.~(\ref{eqA11}) into Eq.~(\ref{eq19}) and performs numerical calculation straightforwardly, $\tilde{r}_{\textrm{dGL}}$ becomes $5.7\times 10^{-4}$, which is %nearly equal to the value of 
in reasonable agreement with $r_{\textrm{dGL}}$. %in the previous paragraph. 
Moreover, it can be deduced from Eq.~(\ref{eqA12}) that as long as the $q^3$-order term in $\alpha_1$ is not to be evaluated, the SD effect would not be expected if $\alpha_3$ and $\beta_3$ are neglected. In this way, the SD effect based on the usual GL theory up to the $q^2$-order stems from the appearance of $\alpha_3$ and $\beta_3$ in the GL free energy, at least when the condition $\varphi_1\gg\varphi_2$ is satisfied.

Although the SC effect occurs in this approach, the diode quality factor $r_{\rm dGL}$ is tiny and does not coincide with the result in the BdG mean-field theory. As we see below, in the BdG mean-field framework, $r_{\rm d}$ is negative in a low magnetic field and high temperature region. This discrepancy arises from neglecting some higher order terms in $q$ and $h$, such as $qh$-order terms in $\beta_i$ and $q^3$-order terms in $\alpha_i$. In other words, the higher-order derivative terms play a qualitatively essential role in the SD effect. It is expected that the GL theory is quantitatively consistent with the BdG theory near the transition temperature when the higher-order derivative terms are appropriately taken into account. However, we will not go any further into the GL framework and move to the microscopic calculation in the BdG framework.

\section{BdG Mean-field theory}
\label{sec:BdG}

\subsection{Gap equations}
\label{sec:gap_equation}

The first step to obtaining the mean-field free energy is to solve the saddle point equations of a given effective action for order parameters. In the model worked on, this can be achieved by considering the equation $\partial S_{\mathrm{eff}}/\partial \bar{\varphi}_i=0$, where $S_{\mathrm{eff}}$ is given by Eq.~(\ref{eq17}). Such a procedure brings us to a couple of gap equations as follows:
\begin{widetext}
\begin{align*}
 \varphi_i(\bm{q}) =  \frac{\lambda_i T }{2} \Sum{\bm{k},\omega_n}{} \, & \Biggr\{  \frac{  \lra{\varphi_1(\bm{q}) \Psi_{1\da\ua\bm{k}} + \varphi_2(\bm{q}) \Psi_{2\da\ua\bm{k}} } \Psi_{i\da\ua\bm{k}}  }{  (-i\omega_n + \eps_{\bm{k}_+} \!\! + h)(i\omega_n + \eps_{-\bm{k}_-} \!\!- h)  + | \varphi_1(\bm{q}) \Psi_{1\da\ua\bm{k}} + \varphi_2(\bm{q}) \Psi_{2\da\ua\bm{k}}    |^2             } \\
& \quad~~ + \frac{  \lra{\varphi_1(\bm{q}) \Psi_{1\ua\da\bm{k}} + \varphi_2(\bm{q}) \Psi_{2\ua\da\bm{k}} } \Psi_{i\ua\da\bm{k}}  }{  (-i\omega_n + \eps_{\bm{k}_+} \!\! - h)(i\omega_n + \eps_{-\bm{k}_-} \!\! + h)  + | \varphi_1(\bm{q}) \Psi_{1\ua\da\bm{k}} + \varphi_2(\bm{q}) \Psi_{2\ua\da\bm{k}}    |^2             } \Biggr\}, \tag{24} \label{eq24}
\end{align*}
where $i=1,2$. Now, the mean-field free energy $F(q)$ of the superconducting state with Cooper-pair momentum $q$ can be calculated by solving the above equations for $\varphi_i$ and substituting them into the expression of free energy, which is given below:
\[    F(q) = -\frac{T}{2}\sideset{}{'}\sum_{\bm{k}}\sum_{\omega_n} \mathrm{Tr}\, \ln \mathcal{G}^{-1}(\omega_n,q,\bm{k}) + \frac{T}{2} \bigg[\sideset{}{'}\sum_{\bm{k}}\sum_{\omega_n} \mathrm{Tr}\, \ln \mathcal{G}^{-1}(\omega_n,q,\bm{k})\bigg]_{\varphi_i=0} + \sum_i \frac{1}{\lambda_i}|\varphi_i(q)|^2.    \tag{25}   \label{eq25}         \]
\end{widetext}
The second term in the right-hand side of the above equation is responsible for the free energy of the normal state. More explicit expressions of Eqs.~(\ref{eq24}) and (\ref{eq25}) can be found in Appendix C. Like in the GL framework, it can be shown that the internal Josephson phase $\phi$ satisfying the condition $\cos\phi=\pm1$ is a trivial stationary point of $F(q)$. A nontrivial one is also possible. However, unlike the GL framework, finding such a nontrivial solution is difficult in the present framework since one has to minimize $F(q)$ at each of $\phi\in[0,\pi]$ numerically. Thus, here we assume that $F(q)$ is minimized when the Josephson phase $\phi$ satisfies the condition $\cos\phi=\pm1$, in which the value of $\phi$ is restricted to $\phi=0$ or $\pi$ as in the GL theory (Sec.~\ref{sec:numerical_GL}). For convenience, let us extend the domain of $\varphi_2$ from positive real numbers to real numbers. In this way, the sign of $\varphi_2$ will reflect the Josephson phase directly. The supercurrent is calculated in the same way as done in Sec.~\ref{sec:GL}. 

The physical parameters necessary for the numerical calculation, $gN(0),~v_a N(0),~v_mN(0),~\mu/T_{{\rm c}0}$, and $\omega_{\rm c}/\mu$, are given as the same as in Sec.~\ref{sec:numerical_GL}.  However, the integration over momentum is now limited to the bounded interval $[\Lambda_-,\Lambda_+]$. As aforementioned in Sec~\ref{sec:model}, the interval is fixed here independent of $q$ in the same way as the BCS theory. Accordingly, $\Lambda_{\pm}$ is given by $\Lambda_{\pm}/p_{\rm F} = (1 \pm \omega_{\rm c}/\mu)^{1/2}$. In this way, the reduced dimensionless mean-field free energy $f_{\mathrm{mf}} \equiv F(N(0)T^2_{{\rm c}0})^{-1}$ can be calculated as a function of $q/p_{\rm F}$, $t$, and $\tilde{h}$.

\subsection{Numerical results}
\label{sec:numerical_results}

\captionsetup{font=normal,justification=raggedright,singlelinecheck=false}

\begin{figure}[tbp]

\begin{minipage}{\hsize}

\subfloat[$q_0$\label{hta}]{\includegraphics[width=8.1cm]{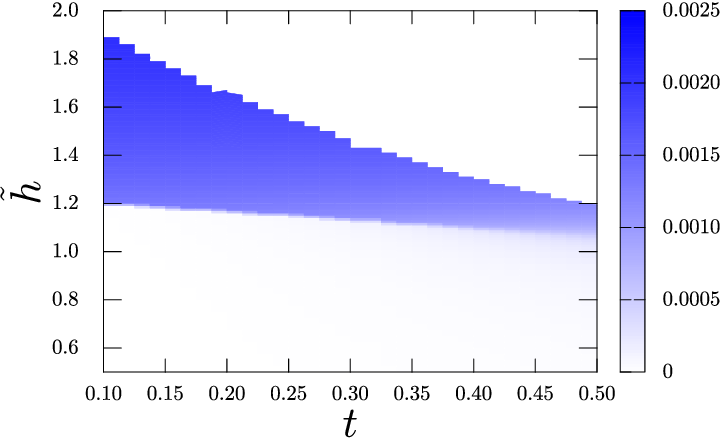}}
\end{minipage}

\begin{minipage}{\hsize}
\subfloat[$\varphi_2$\label{htb}]{\includegraphics[width=8.1cm]{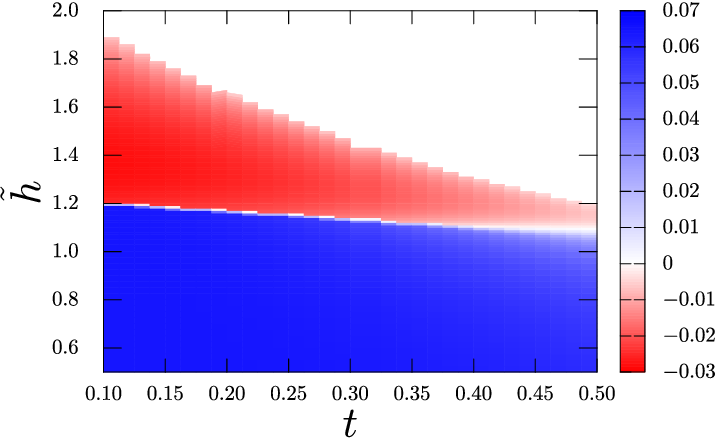}}
\end{minipage}

\begin{minipage}{\hsize}
\subfloat[$r_{\rm d}$\label{htc}]{\includegraphics[width=8.1cm]{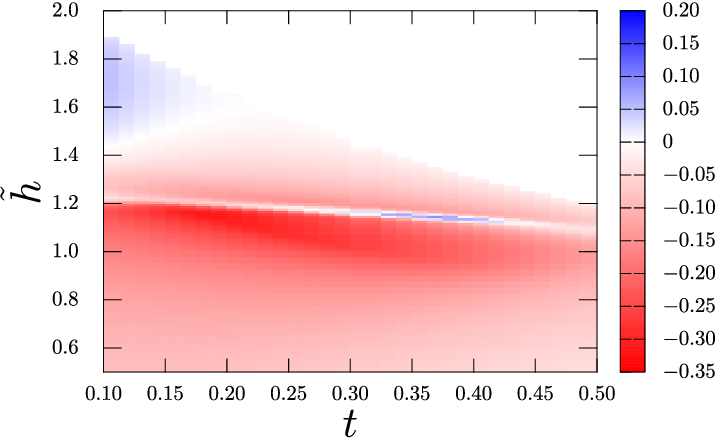}}
\end{minipage}

\caption{Color maps of the $\tilde{h}$-$t$ phase diagram. The color gradation indicates (a) the Cooper-pair momentum $q_0$ in the most stable state, (b) an order parameter $\varphi_2$, and (c) the diode quality factor $r_{\rm d}$. 
In (b), $\varphi_2$ is defined as a real number to reflect the Josephson phase. See Sec.~\ref{sec:gap_equation} for further details. In (c), the white band lying between approximately $\tilde{h}=1.1$ and 1.3 is referred to as ``region E" in the text.}

\label{ht}
\end{figure}

\begin{figure}[t]

\centering
\includegraphics[width=7.2cm]{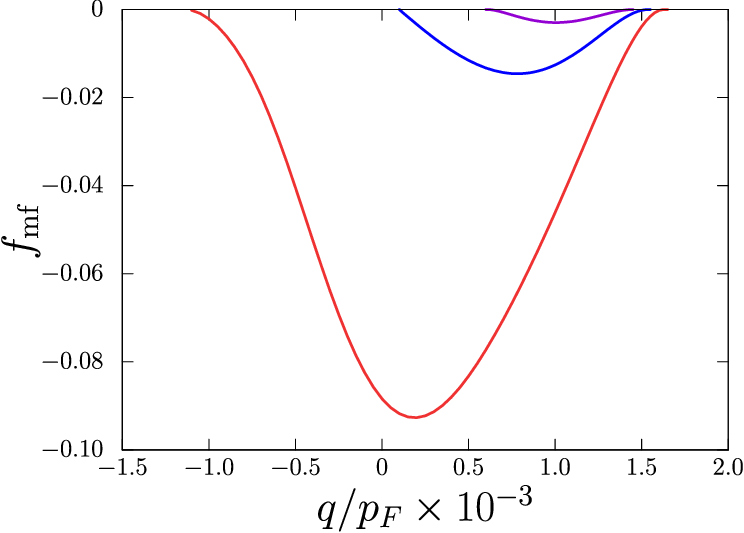}

\caption{Cooper-pair momentum dependence of the reduced dimensionless mean-field free energy $f_{\mathrm{mf}}$ obtained in the BdG theory. The temperature is $t=0.5$, and various high magnetic fields $\tilde{h}=1.15,~1.10$, and $1.00$ are adopted from the top to bottom.}

\label{t0.5}
\end{figure}

\begin{figure*}[t]

\begin{minipage}{0.49\hsize}
\subfloat[$\tilde{h}=0.90$\label{t0.1a}]{\includegraphics[width=7.5cm]{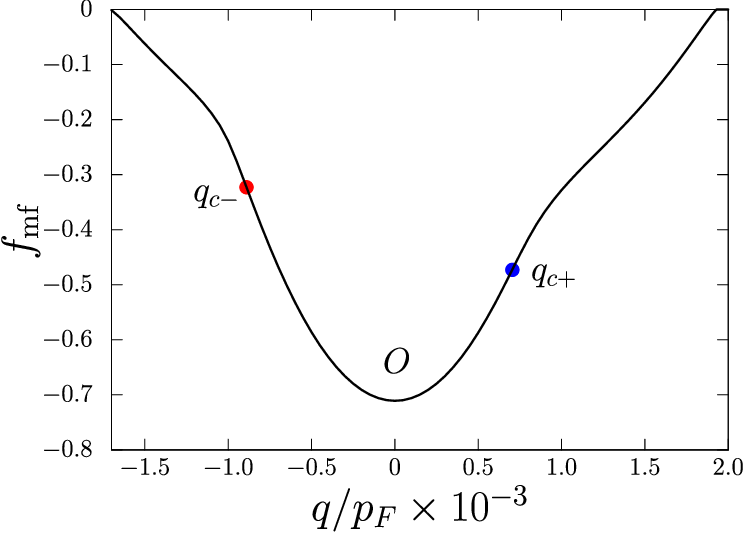}}
\end{minipage}
\begin{minipage}{0.49\hsize}
\subfloat[$\tilde{h}=1.15$\label{t0.1b}]{\includegraphics[width=7.5cm]{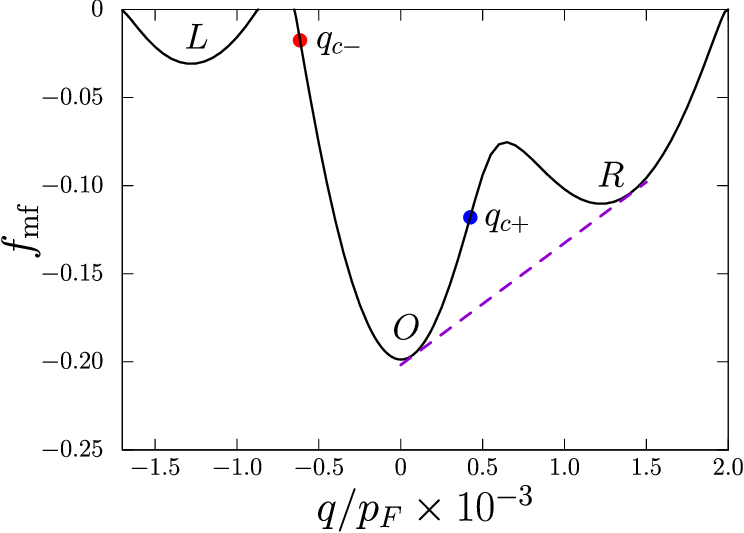}}
\end{minipage}

\begin{minipage}{0.49\hsize}
\subfloat[$\tilde{h}=1.20$\label{t0.1c}]{\includegraphics[width=7.5cm]{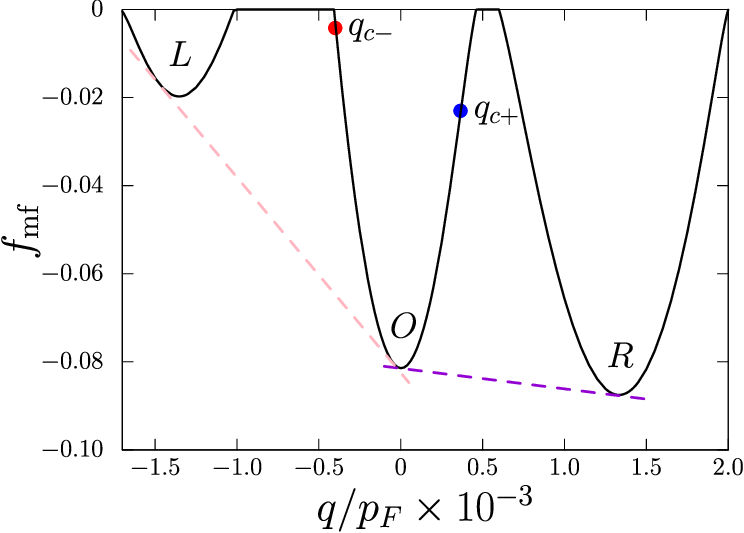}}
\end{minipage}
\begin{minipage}{0.49\hsize}
\subfloat[$\tilde{h}=1.25$\label{t0.1d}]{\includegraphics[width=7.5cm]{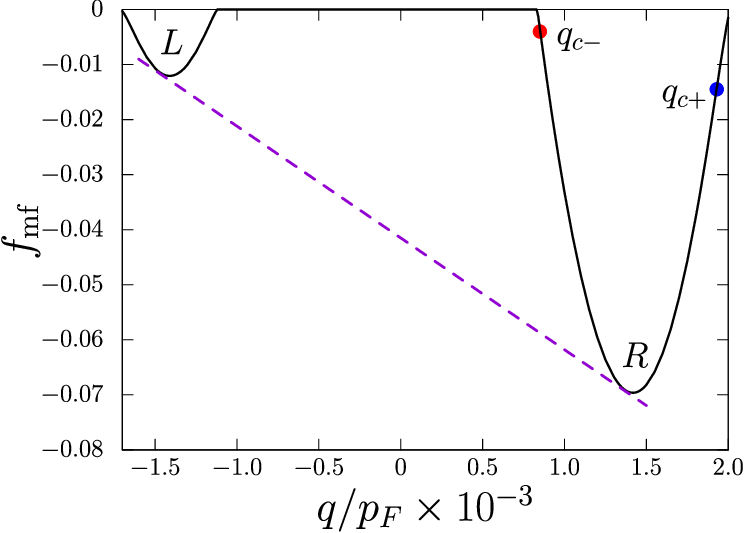}}
\end{minipage}

\caption{Cooper-pair momentum dependence of the reduced dimensionless mean-field free energy $f_{\mathrm{mf}}$ at the temperature $t=0.1$ and various high magnetic fields $\tilde{h}$. Dashed lines in the figure show tangent lines. The red and blue dots denote the Cooper-pair momentum corresponding to the critical current in the $-z$-direction and $+z$-direction, respectively.}
\label{t0.1}
\end{figure*}

First, we describe some features of $\tilde{h}$-$t$ phase diagrams of the model when the supercurrent $I$ is 0. Let $q_0$ be the Cooper-pair momentum at the lowest free energy. As shown in the phase diagram in Fig.~\ref{hta}, there is a transition of $q_0$ in the relatively high magnetic field region. While a crossover of $q_0$ appears at high temperatures, it turns into a first-order transition at sufficiently low temperatures. The critical point or the endpoint of the first-order transition at the high-temperature side is approximately located at the field $\tilde{h}=1.08\pm0.02$ and temperature $t=0.425\pm0.025$. It must also be emphasized that the first-order transition is due to the growth of metastable states accompanied by the mean-field theory. While metastable states do not exist in the crossover regime, they are manifest in the first-order transition regime. The change of the mean-field free energy upon increasing the external magnetic field is shown at $t=0.5$, which is in the crossover regime, and at $t=0.1$, which is the first-order transition regime, in Figs.~\ref{t0.5} and \ref{t0.1}, respectively. %Based on these results, we conclude 
These results reveal that the first-order transition at low temperatures is caused by the manifestation of metastable states, and the two states with different Cooper-pair momentum $q_0$ should coexist at the first-order transition. These features are the same as the case of a single-component superconductor~\cite{Daido1, Daido2}. 

\captionsetup{font=normal,justification=raggedright,singlelinecheck=false}

A distinctive feature of the model can be found by looking at the Josephson phase $\phi$. As shown in Fig.~\ref{htb}, the order parameter $\varphi_2$ changes the sign with the transition of $q_0$. This means that the Josephson phase discontinuously changes from $0$ to $\pi$ at the first-order transition or crossover of $q_0$ upon increasing the external magnetic field at a fixed temperature. In accordance with the nature of the phase transition, the change in the value of $\varphi_2$ is, as well as $q_0$, discontinuous at low temperatures while continuous at high temperatures. As a result of the latter, at high temperatures where $\varphi_2$ changes continuously, there is a line where the Josephson phase cannot be defined because $\varphi_2$ becomes zero around the crossover line of $q_0$. Furthermore, the discontinuous change in $\varphi_2$ at low temperatures implies that the phases coexisting at the first-order transition are different not only in the momentum $q_0$ but also in the Josephson phase $\phi$.

Let us move to the case when the system is under a constant supercurrent. We show a color map of the diode quality factor varying with external magnetic fields and temperatures in Fig.~\ref{htc}. Below $\tilde{h} \sim 1.1$, the signs of the diode quality factors are all negative. In contrast, at high enough magnetic fields and low enough temperatures, for example, at $\tilde{h}\gtrsim 1.4$ and $t\lesssim0.1$, the sign of the diode quality factor changes from negative to positive. This feature coincides qualitatively with the results in a single-component superconductor~\cite{Daido1}. However, in the range of relatively high magnetic fields $1.1<\tilde{h}<1.4$, a novel behavior of the SD effect appears. There is a region where the diode quality factor takes a local minimum value with a negative sign or a local maximum value with a positive sign. Such a region is located slightly above the transition line of $q_0$. We term it ``region E". The presence of this region differs from the result of single-component superconductors~\cite{Daido1, Daido2, Ilic}, where single or double sign reversal occurs at all temperatures in the corresponding region. Thus, the region E is a hallmark of the SD effect arising from the DM interaction. Slightly above the region E, the diode quality factor reaches a local maximum with a negative sign. The diode quality factor reaches the global maximum with a negative sign below the region E and slightly below the transition line of $q_0$. 

\begin{figure}[t]

\captionsetup{font=normal,justification=raggedright,singlelinecheck=false}

\begin{minipage}{\hsize}
\subfloat[$t=0.1$\label{rha}]{\includegraphics[width=7.5cm]{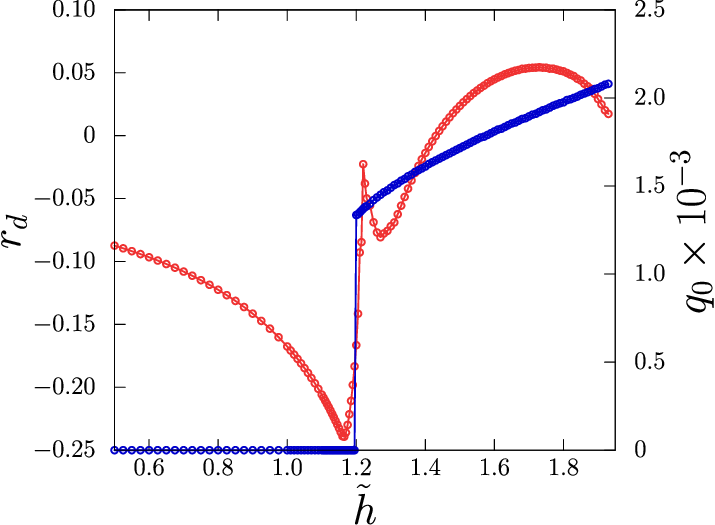}}
\end{minipage}

\begin{minipage}{\hsize}
\subfloat[$t=0.5$\label{rhb}]{\includegraphics[width=7.5cm]{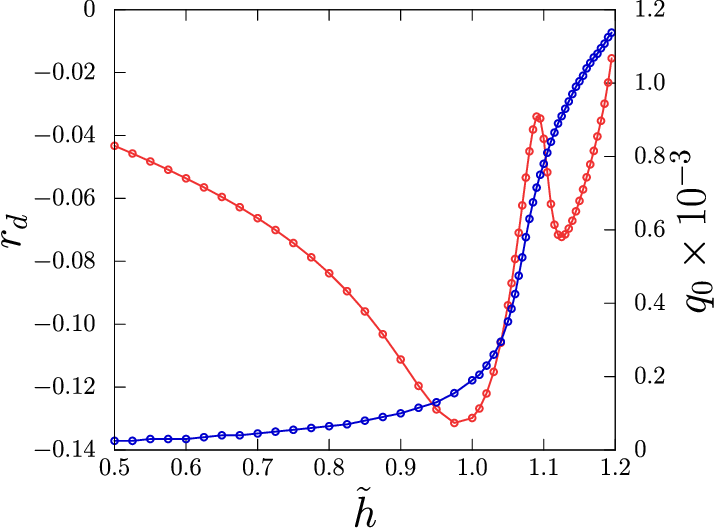}}
\end{minipage}

\caption{Magnetic field dependence of the diode quality factors $r_{\rm d}$ and the most stable Cooper-pair momentum $q_0$ at (a) $t=0.1$ and (b) $t=0.5$. The red (blue) line denotes the $r_d$ ($q_0$) curve. At $t=0.1$, the first-order transition of $q_0$ occurs at $\tilde{h}=1.2$, while the local minimum in the magnitude of $r_{\rm d}$ in the region E locates at $\tilde{h}=1.22$. At $t=0.5$, $q_0$ and $r_{\rm d}$ vary smoothly over the range of magnetic fields, and the $r_{\rm d}$ curve reaches its local minimum in magnitude in the region E at $\tilde{h}=1.09$. }

\label{rh}
\end{figure}

Readers might wonder why the region E and the first-order transition of $q_0$ do not perfectly overlap, as seen in Figs.~\ref{hta} and \ref{htc}, or much easier in Fig.~\ref{rha}. A reason for this can be grasped by paying attention to the relation between the growth of metastable valleys and values of $q_{{\rm c}\pm}$ at which the supercurrent reaches a critical current. As shown in Figs.~\ref{t0.1b} and \ref{t0.1c}, upon increasing $\tilde{h}=1.15$ to $\tilde{h}=1.20$ at $t=0.1$, the minimum of free energy $f_{\mathrm{mf}}$ changes from the valley $O$ to the valley $R$, while the place of $q_{{\rm c}\pm}$ is still in the same valley $O$. As a result, in the vicinity of the first-order transition, $q_{{\rm c}\pm}$ and $I_{{\rm c}\pm}$, in other words, the diode quality factor $r_{\rm d}$ can still vary smoothly. In this way, the region E in the low temperature region and the first-order transition of $q_0$ do not completely overlap.     

From Fig.~\ref{rh}, we further find a peculiar behavior of the diode quality factor in the region E. Upon increasing the magnetic field at $t=0.1$, it can be seen from Fig.~\ref{rha} that the diode quality factor curve reaches a local minimum in magnitude and becomes non-differentiable at a certain magnetic field. Indeed, following our numerical data, such a behavior occurs at the same time as the Cooper-pair momenta giving critical currents to move from the valley $O$ to the valley $R$. On the other hand, in Fig.~\ref{rhb} for $t=0.5$, at which metastable valleys are absent, the diode quality factor curve varies smoothly over the range of magnetic fields. The values of the magnetic field and temperature at which the curve becomes non-differentiable are vague at this stage. However, our numerical data affirms that the curves become sharper at the extreme points in region E upon cooling, and we argue that the growth of metastable valleys in the low temperature regime causes \YY{the non-differentiable} behavior.

\subsection{Phase coexistence under a steady current}
\label{sec:phase_coexistence}

In our theoretical framework, metastable states appear in the high magnetic field region above the transition line of $q_0$. Furthermore, as shown in Fig.~\ref{t0.1c}, the critical currents can be realized in a valley of the metastable state. A question arises of how this can happen if starting in a stable state such as the valley $R$ in Fig.~\ref{t0.1c}. We answer a question through the following discussion. Suppose a stable state is reached in the region above the transition line of $q_0$. We then apply an external current to the system slowly. As a result, at a certain current $I_{\rm M}$ with corresponding momentum $q_R$, the condensation energy $F(q)$ decreases to the value that the Gibbs free energy $G(I_{\mathrm{M}})$ in the valley of the stable state becomes equal to that in another valley with corresponding momentum $q_O$. This current $I_{\rm M}$ can be obtained by drawing a tangent line connecting the point of each valley. Examples are shown in Figs.~\ref{t0.1b}, \ref{t0.1c}, and \ref{t0.1d}. The number of points from each valley that shares the same tangent line can reach at most three at some external magnetic fields and temperatures within this model. Points on such a tangent line share the same Gibbs free energy $G(I_{\rm M})$. Consequently, phase coexistence should occur between the states of distinct valleys, as in the phase coexistence of liquid and gas \cite{Chaikin}. For concreteness, we consider the case where two phases coexist, as in Fig.~\ref{t0.1c}. Then, $I_{\rm M}$ is given in terms of $F(q_O)$, $F(q_R)$, $q_O$, and $q_R$ as follows:
\[       I_{\rm M} =   \frac{F(q_O) - F(q_R)}{q_O-q_R}.  \tag{26}   \]
Let $\chi$ and $1-\chi$ be the volume fraction of $q_O$- and $q_R$-states, respectively. Thus, the condensation energy of each point on the tangent line, $f_{\mathrm{s}}(q)$, is
\[     f_{\mathrm{s}}(q) = \chi F(q_O) + (1-\chi) F(q_R).     \tag{27}          \]

Further increasing current beyond $|I_{\rm M}|$ slowly, the state of the whole system changes its Cooper-pair momentum discontinuously from a value of the valley $R$ to that of the valley $O$. This means that the states on the thick black curve between two tangent points, namely $q_O$ and $q_R$, are \textit{prohibited} when the system settles down in the state whose Gibbs free energy becomes the minimum. Again, at some currents $|I|>|I_{\rm M}|$ with corresponding momentum in the valley $O$, it turns out that the system can go to the valley $L$ through another tangent line. If this is the case, the critical current in the negative $z$-direction can be different from the value given in Eq.~(\ref{eq22}), and the diode quality factor also changes. The diode quality factor defined by Eq.~(\ref{eq22}) is $r_{\rm d}=-0.23$, $-0.17$, and $-0.07$ in Figs.~\ref{t0.1b}, \ref{t0.1c}, and \ref{t0.1d}, while the above thermodynamic argument gives $-0.36$, $0.43$, and $0.51$, respectively. Not only the magnitude but also the sign changes when the thermodynamic argument applies. To prevent this, we consider how a supercooled liquid is realized. Such a phenomenon occurs by a quick drop in temperature towards a freezing temperature \cite{Debenedetti}. As an analogy of this phenomenon, we believe that the system can reach the state with momentum $q_{{\rm c}-}$ in the valley $O$ by rapidly increasing current. By such a rapid change, the state with $q_{{\rm c}+}$ in the valley $O$ can also be reached. However, these states that do not give the minimum to $G(I)$ for each $I$ could remain only for a finite time. Let $I'_{{\rm c}\pm}\geq 0$ be the critical current calculated from the minimum of $G(I)$ in the $\pm z$-direction. Our theoretical argument implies intriguing dynamics, such as relaxation to normal state or other superconducting states, leading to the inevitable transient SD effect if the current $I$ satisfying $|I_{c\pm}|>|I|>|I'_{c\pm}|$ is applied and the superconducting state with corresponding Cooper-pair momentum $q$ realizes.

Note that the above discussion is expected to be general and can also be applied to single-component superconductors. However, here we have not considered surface energies between coexisting phases. The dynamics of the superconducting state are known to depend on the setup and are desired to be clarified for further understanding of the SD effect.

\section{Comments on purely 2D Systems}

So far, we have studied the SD effect at the mean-field level with effectively three-dimensional superconducting thin films in mind. Here, let us point out two issues related to a purely 2D system, which can be realized in atomically thin films, to stress the fundamental differences between quasi-2D and purely 2D superconductors. The following comments are important for those who wish to study our model in the context of purely 2D systems or to seek out the SD effect in purely 2D systems.

Firstly, in a 2D system, even under an in-plane magnetic field, the superconducting transition should be the BKT transition, in which the effect of the in-plane Zeeman field is renormalized to the superfluid weight \cite{Hoshino, Julku, Kitamura}. Using the Nelson-Kosterlitz (NK) criterion \cite{Nelson}, we can estimate the BKT transition temperature $T_{\mathrm{BKT}}$ at various magnetic fields. To do this, we have to calculate the superfluid weight matrix $\hat{\rho}_w$, whose $ij$-component is given by $(\hat{\rho}_{w})_{ij} \propto [ \partial^2 F(\bm{A})/\partial A_i \partial A_j ]_{\bm{A}=0}$, where $\bm{A}$ is a virtual gauge field \cite{Kukja}. This can further be simplified as follows:
\[   (\hat{\rho}_w)_{ij} \propto \frac{\partial^2}{\partial q_i \partial q_j} F(\bm{q}). \tag{28}  \label{eq28}   \]
Therefore, the superfluid weight is the second-order derivative of the condensation energy with respect to $q$. Using the above equation, the NK criterion can be written as below \cite{Julku}: 
\[        \lra{\mathrm{det}\,\hat{\rho}_w}^{1/2} \approx \frac{2}{\pi}  T_{\mathrm{BKT}}.          \tag{29} \]
The approximation notation is used above because $\rho_w$ in Eq.~(\ref{eq28}) has not yet been renormalized by thermally-induced vortex-antivortex pairs. 

In applying the BKT theory to a two-component model like the model studied here, although the complete theory of such systems has still not been established, there is a simulation study of a 2D system of two-component mixed Bose gases~\cite{Kobayashi} that provides useful wisdom. There, it is shown that in a case with nonzero Josephson coupling, the NK criterion is still valid. Therefore, it would be inclined to conclude that the BKT theory can be applied to the model studied here. However, there is a difference between the model studied in Ref.~\cite{Kobayashi} and our model. The Josephson couplings in quaternary terms of order parameters had not been included in the former but were included in the latter, as can be seen in Eq.~(\ref{eq19}). Whether it is significant is an open question.

Secondly, vortex-antivortex unbinding under a finite electric current would result in the \textit{absence} of the SD effect, accompanied by a discontinuous change in resistance between zero ohms state and finite ohms state, in a purely 2D system. In purely 2D superconductors, according to Refs.~\cite{Halperin} and \cite{Kadin}, although the in-plane DC resistivity is strictly zero only in the limit of vanishing current, the current-voltage characteristic becomes nonlinear in a \textit{not-too-low} current region. Indeed, a true 2D system is regarded as a superconductor in this sense. However, lacking P and T symmetry in such a system can cause the directional asymmetry of the nonlinear current-voltage characteristic under a low electric current. A detailed study of them will be presented elsewhere.

\section{Summary and Discussions}

In this paper, we have studied the superconducting diode effect in a two-component superconductor with the Zeeman coupling and weak DM interaction based on the GL and BdG mean-field framework. In the GL framework, it has become apparent that calculating each GL coefficient up to $q^2$-order is sufficient to demonstrate the SD effect. However, a numerical result of the model showed that even in the vicinity of the zero-field transition temperature of superconductivity, the diode quality factor is tiny and different from the BdG framework due to neglecting higher-order terms of $q$ in the GL coefficients. For the BdG mean-field framework, the internal Josephson coupling changes its sign in a low temperature and high magnetic field region in thermal equilibrium. The first-order transition of Cooper-pair momentum also appears in a low temperature region. At the transition, the phases with different Cooper-pair momentum and Josephson phase coexist. Under the supercurrent, the diode quality factor calculated from the extremes of supercurrent changes the sign in two regions: One is located in a low temperature and high magnetic field region, and the other is in a finite range of temperatures named region E, located at moderately high magnetic fields. In region E, the diode quality factor calculated from the extremes of the supercurrent shows non-monotonic magnetic field dependence and changes its sign in a high temperature region but does not change at sufficiently low temperatures.

Furthermore, we went deep into what the critical current would imply to the stability of superconductivity. Interpreting a supercurrent as a control variable and using thermodynamic arguments, we argued that the corresponding Cooper-pair momentum of the system under a constant supercurrent is a value such that the counterpart Gibbs free energy, being a function of supercurrent, reaches its minimum. Consequently, phase coexistence and first-order transition of Cooper-pair momentum upon increasing the external current are predicted in a low temperature and high field region. A critical current and diode quality factor can also differ from those calculated from the extremes of the supercurrent. The critical current predicted from the global extremum of supercurrents may be realized by a rapid change of supercurrent as an analogy of the supercooled liquid if we assume a supercurrent plays a role in temperature. However, a state achieved by such a procedure should not last an infinite time if the thermodynamic argument for the supercurrent applies to the setup of the system, since it is not a truly stable SC state in stationary thermodynamic equilibrium. In this way, we conclude that the critical current significantly depends on the dynamics when multiple metastable states exist in a low temperature and high field regime, and attention should be paid to the definition of the diode quality factor. In other words, the transient dynamics of the SD effect may occur associated with the transition of helical superconducting states. It can be a signature of the helical superconductivity in noncentrosymmetric superconductors.

In this paper, only the FF state, being spatially homogeneous, was discussed. For the Larkin-Ovchinnikov (LO) state~\cite{Larkin}, our approach starting from a momentum-space representation of the action cannot give the same result as one expects from a real-space representation. The problem stems from the fact that it is not easy to perform the Fourier transform of the action from a real-space representation to a momentum-space representation. For example,  for simplicity, let us consider a single-component system in which the gap function in the real-space representation is $\Delta(x)=A\exp\lr{iq_1x}+B\exp\lr{iq_2x}~(q_1\not=q_2)$. This gap function realizes an FFLO state \cite{Matsuda}. If $q_1=-q_2$ but $A\not=B$, the state is known as a stripe state \cite{Agterberg}. Thus, one has $\mathcal{G}^{-1}(\tau,x)$ in the form that $\hat{\Delta}_{\pm}$ in Eq.~(\ref{16a}) is replaced by two matrices with one proportional to $\Delta(x)$ and other one is proportional to $\Delta^*(x)$. Now, it can be seen that the factors $\exp\lr{\pm iq_ix}$ in $\mathcal{G}^{-1}(\tau,x)$ cannot be canceled by a unitary transformation, meaning that one cannot reach a simple form of $\mathcal{G}^{-1}(\tau,k)$ like Eq.~(\ref{16a}). In this way, our approach cannot straightforwardly be generalized to the case with a single phase composed of multiple Cooer-pair momentum $\bm{q}$ giving rise to spatial inhomogeneity. Starting from a real-space representation of the action of our model is also challenging. Nevertheless, from the centrosymmetry of the LO state, one can quickly conclude that the SD effect should not occur, at least if the external current changes adiabatically. It could be interesting to study the SD effect in FFLO states with spacial modulation because it appears in a phase diagram of the equilibrium state~\cite{NCSC_book, Smidman}. 

\section*{Acknowledgments}
We are grateful to R. Ikeda and A. Daido for fruitful discussions.
This work was supported by JSPS KAKENHI (Grant Nos.~JP21K18145, JP22H01181, JP22H04933, JP23K17353).

\appendix*

\def\thesection{\Alph{section}}

\section{GL coefficients}

The GL coefficients of the two-component model: in Eq.~\eqref{eq19} are obtained as follows:
\begin{align*}
\alpha_1 &=\frac{1}{\lambda_+} - \frac{1}{4}N(0)\lr{3 + \cos2\chi }\lra{\log\frac{2\eps_c}{\pi T} +\gamma } \\
&\quad -2\beta_0 \sin2\chi \, \frac{hv_Fq}{T^2} + \frac{1}{8}\beta_0 \lr{3+\cos^2\chi } \frac{v^2_Fq^2}{T^2} ,  \tag{A1}  \\
\alpha_2 &=\frac{1}{\lambda_-} - \frac{1}{4}N(0)\lr{3 - \cos2\chi}\lra{\log\frac{2\eps_c}{\pi T} +\gamma } \\
&\quad +2\beta_0 \sin2\chi \, \frac{hv_Fq}{T^2} + \frac{1}{8}\beta_0 \lr{3+\sin^2\chi } \frac{v^2_Fq^2}{T^2},  \tag{A2} \\
\alpha_3 &=-\frac{1}{4} N(0)\sin2\chi \lra{\log\frac{2\eps_c}{\pi T} +\gamma } \\
&\quad + 2\beta_0 \cos2\chi \, \frac{hv_Fq}{T^2} +  \frac{1}{16} \beta_0 \sin2\chi \, \frac{v^2_Fq^2}{T^2} , \tag{A3}
\end{align*}

\begin{align*}
\beta_1 &= \frac{1}{T^2}\beta_0 \lr{1 + \sin^2\chi - \frac{13}{8} \sin^4\chi},   \tag{A4}  \\
\beta_2 &= \frac{1}{T^2}\beta_0 \lr{1 + \cos^2\chi - \frac{13}{8} \cos^4\chi},   \tag{A5}  \\
\beta_3 &= -\frac{1}{2T^2}\beta_0 \sin2\chi \lr{1 - \frac{13}{4} \sin^2\chi},        \tag{A6} \\
\beta_4 &= -\frac{1}{2T^2}\beta_0 \sin2\chi \lr{1 - \frac{13}{4} \cos^2\chi},        \tag{A7} \\
\beta_5 &= 2\beta_6 = \frac{1}{T^2}\beta_0 \lr{ 5- \frac{13}{16} \sin^22\chi },    \tag{A8}
\end{align*}
where $\beta_0 =7\zeta(3)N(0)/(16\pi^2)$, $N(0) = mA/(2\pi)$, and $A$ denotes system's area. 

In a one-component case with $\lambda_-\simeq 0$,  the GL free energy can be written as $F_{\mathrm{GL}} = \alpha_1\varphi^2_1 + \beta_1\varphi_1^4$. Although an explicit numerical calculation for such a case was not performed in this paper, we give here the GL coefficients as follows:

\begin{align*}
\alpha_1 &=\frac{1}{\lambda_+} - \frac{1}{4}N(0)\lr{3 + \cos2\chi }\lra{\log\frac{2\eps_c}{\pi T} +\gamma } \\
&\quad\quad -2\beta_0 \sin2\chi \, \frac{hv_Fq}{T^2} + \frac{1}{8}\beta_0 \lr{3+\cos^2\chi } \frac{v^2_Fq^2}{T^2} \\
&\quad\quad + \frac{93  \, \xi(5)}{112\, \pi^2 \xi(3)} \beta_0 \sin2\chi \frac{h \, (v_Fq)^3}{T^4},  \tag{A9}  \\
\beta_1 &= \frac{1}{T^2}\beta_0 \lr{1 + \sin^2\chi - \frac{13}{8} \sin^4\chi} \\
&\quad\quad + \frac{589  \, \xi(5)}{448\, \pi^2 \xi(3)} \beta_0 \sin2\chi\, \lr{3+\cos^2x} \frac{h v_Fq}{T^2}.   \tag{A10}  
\end{align*}

\section{Asymptotic solutions for the GL equations}

Following the procedures described in the main text, we can evaluate $\varphi^2_1$ and $F_{\mathrm{GL}}$ as follows:
{\[  \varphi^2_1 = -\frac{\alpha_1}{2\beta_1}\lra{1+\frac{3\alpha_3\beta_3}{2\alpha_2\beta_1}} + \frac{\alpha^2_3}{2\alpha_2\beta_1} + O\lr{\alpha^2_3/\alpha^2_2},  \tag{A11} \label{eqA11} \] 
\[ F_{\mathrm{GL}} = -\frac{\alpha^2_1}{4\beta_1}\lra{1 + \frac{2\alpha_3\beta_3}{\alpha_2\beta_1}} + O\lr{\alpha^2_3/\alpha^2_2}.  \tag{A12} \label{eqA12} \]}

\begin{widetext}

\section{Expressions of Eqs.~(\ref{eq24}) and (\ref{eq25})}

{After taking summation of Matsubara frequency in Eqs.~(\ref{eq24}) and (\ref{eq25}), one can obtain the following:
\[    \varphi_i(q) = \frac{\lambda_i T}{2} \sideset{}{'}\sum_{\bm{k}} \lra{ \tanh \frac{\eps_+(\bm{k},q)}{2T} + \tanh \frac{\eps_-(\bm{k},q)}{2T} }  \times \frac{\Delta_G(q,\theta)}{E_G(\bm{k},q)}\Delta_i(\theta),   \tag{A13} \label{eqA13}  \]
\[ F_s = - \sideset{}{'}\sum_{\bm{k}} \Big[ T \ln (1+e^{\eps_+(\bm{k},q)/T})(1+e^{\eps_-(\bm{k},q)/T})   + E_G(\bm{k},q)  \Big]  + \sum_i \frac{1}{\lambda_i}|\varphi_i(q)|^2,  \YY{ \tag{A14} }   \label{eqA14}  \]}
\end{widetext}
where $F_s$ is the free energy of the superconducting part. To obtain Eq. (\ref{eq25}), one has to subtract from Eq. (\ref{eqA14}) the normal part, which is calculated by substituting $\varphi_i=0$ into Eq. (\ref{eqA14}). Moreover, the other variables are given as follows:
\begin{align*}
\Delta_G(q,\theta)& = \varphi_1(q)\Delta_1(\theta) + \varphi_2(q)\Delta_2(\theta),   \tag{A15} \\
\Delta_1(\theta) & = -\cos\chi + \sin\chi\cos\theta, \tag{A16}     \\
\Delta_2(\theta)& = -\sin\chi - \cos\chi\cos\theta,    \tag{A17}  \\
\eps_{\pm}(\bm{k},q) &= E_G(\bm{k},q) \pm \bigg( \frac{kq}{2m} \cos\theta +h \bigg),  \tag{A18}  
\end{align*}
and
\[  E_G(\bm{k},q) = \lra{ \lr{\frac{k^2}{2m} + \frac{q^2}{8m} -\mu}^2 + [\Delta_G(\theta)]^2 }^{1/2}   \tag{A19}     \]
expresses the quasiparticle's spectrum. In the numerical calculation, we replace the summation notations in Eqs.~(\ref{eqA13}) and (\ref{eqA14}) by
\[  \sideset{}{'}\sum_{\bm{k}}\rightarrow \frac{N(0)}{\pi m} \int_{\Lambda_-}^{\Lambda_+} dk\,k \int_0^{\pi}d\theta, \tag{A20}   \]
where $\Lambda_{\pm}$ is described in the main text.

%$t=0.1,~h=1.15:~f(q)=-69.2q-0.2018$

%Between $q=-0.0001$ and $q=0.0015$,~$t=0.1,~h=1.21:~f(q)=-4.65q-0.0815$.
%Between $q=-0.0015$ and $q=0.0001$,~$t=0.1,~h=1.21:~f(q)=-44.4q-0.0825$.

%$t=0.1,~h=1.25:~f(q)=-20.3q-0.0415$

\end{document}